\documentclass[11pt]{article}
\usepackage{rotating}
\usepackage{pbox}
\usepackage{amsmath}
\usepackage{amssymb}
\usepackage{amsfonts}  
\newcommand{\be}{\begin{equation}}
\newcommand{\ba}{\begin{eqnarray}}
\newcommand{\ee}{\end{equation}}
\newcommand{\ea}{\end{eqnarray}}

\begin{document}

\title{Rationally extended many-body truncated Calogero-Sutherland model}

\author{Rajesh Kumar Yadav$^{a}$\footnote{e-mail address: rajeshastrophysics@gmail.com (R.K.Y)}, Avinash Khare$^{b}$\footnote {e-mail address: khare@physics.univpune.ac.in (A.K)}, Nisha Kumari$^{c}$\footnote{e-mail address: nishaism0086@gmail.com (N.K)} and \\
  Bhabani Prasad Mandal$^{c}$\footnote{e-mail address: bhabani.mandal@gmail.com (B.P.M).}}
 \maketitle
{$~^a$ Department of Physics, S. K. M. University, Dumka-814101, INDIA.\\
 $~^b$Department of Physics, Savitribai Phule Pune University, Pune-411007, INDIA.\\
$~^c$ Department of Physics, Banaras Hindu University, Varanasi-221005, INDIA.}

\begin{abstract}

We construct a rational extension of the truncated Calogero-Sutherland  model 
by Pittman et al. The exact solution of this rationally extended model is 
obtained analytically and it is shown that while the energy eigenvalues remain
unchanged, however the eigenfunctions are completely 
different and written in terms of exceptional $X_1$ Laguerre orthogonal 
polynomials. The rational model is further extended to a more general, 
the $X_m$  case by introducing $m$ dependent interaction
term. As expected, in the special case of $m=0$, the extended model reduces 
to the conventional model of Pittman et al. In the two appropriate limits, 
we thereby obtain rational extensions of the celebrated Calogero-Sutherland
as well as Jain-Khare  models. 
\end{abstract}

\section{Introduction}

The discovery of the two new orthogonal polynomials 
namely the exceptional $X_m$-Laguerre and  exceptional $X_m$-Jacobi orthogonal 
polynomials \cite {dnr1,xm1,xm2} has inspired the discovery of  
a number of new exactly solvable (ES) conventional one-body potentials through
the rational extension of several conventional ES potentials. In most of these
cases while the eigenvalues remain unchanged, the eigenfunctions are in 
terms of these newly discovered exceptional orthogonal polynomials (EOPs) 
\cite{que,bqr,os,hos,hs,gom,qu,yg1,dim,pdm,nfold2,qscat,scatpt,tdse,gtextd, gtextd2, 
para_sym}. The next obvious question is how to construct rational extensions
of many body problems as well as one body non-central but separable potentials. 
Kumari et al, \cite{nrab} took first step in that direction and 
considered the Calogero-Wolfes type three-body problems and constructed a class of corresponding rationally extended 
three-body systems and obtained their exact solutions in terms of $X_m$ 
exceptional Laguerre and $X_m$ exceptional Jacobi polynomials. In another 
paper \cite{nrab_nc} they constructed the rational extension of a number of
non-central but separable one-body potentials and again showed that while
the energy-eigenvalue spectrum is unchanged, the eigenfunctions are now
in terms of $X_m$ Laguerre and $X_m$ Jacobi polynomials. 
The obvious next step is to consider the rational extension of the 
many-body problems. One obvious candidate to consider would be the celebrated 
Calogero-Sutherland (CSM) \cite{cal_71,suth_71} N-body problem on a line with harmonic
confinement. Rational extension of $N$-particle Calogero model with harmonic confining term and arbitrary 
interaction of the form $U(\sqrt{N}\rho)$ ($\rho$ being the radial coordinate), was carried by Basu-Mallick et.al \cite{bpm5}. Bound state eigenfunctions  for specific angular part solution ( hence QES solutions ) were explicitly calculated using supersymmetric technique
 in terms of $X_m$ exceptional Laguerre polynomials. In this context, it is worth recalling 
that several years ago, Jain and Khare (JK) \cite{jain_khare_99} considered a variant of CSM 
on the full line where there was only
nearest and next-to-nearest neighbor interaction through two body and three
body interactions and obtained its eigen spectrum. Recently, Pittman et al 
\cite{pitt_17} generalized the JK model by considering N-body problem on a line
with harmonic confinement in which the tunable inverse square as well as 
the three-body interaction extends over a finite number of neighbors
and were able to obtain its eigenspectrum. One of the nice feature of
this model is that in the appropriate limits its eigen values and 
eigenfunctions smoothly goes over to those of JK and CSM.    
One of the common feature in all the three cases is that a part of the 
eigenfunction is in terms of the celebrated classical Laguerre polynomials. 

The purpose of this note is to consider the rational extension of the 
the truncated Calogero-Sutherland (TCS) model of Pittman et al, \cite {pitt_17} by introducing new
interaction terms over and above the two-body and three-body terms and 
obtain the exact solutions of this model in terms of $X_1$ exceptional 
Laguerre polynomials. We further generalize it to the more general 
$X_m$ Laguerre case by introducing an $m$-dependent polynomial type 
interaction term. It must be mentioned here that the energy eigenvalue
spectrum remains unchanged and is identical to the TCS model. As expected, 
in the special case of $m=0$, the model 
reduces to the usual TCS model \cite{pitt_17}. In the appropriate limits
we thus obtain the rational extension of the celebrated CSM \cite{cal_71,suth_71} as well 
as that of JK model \cite{jain_khare_99}.
 
The plan of the manuscript is as follows:
In section $2$, we briefly recall the TCS model and briefly discuss its 
solutions. In section $3$, we extend the TCS model by introducing a new 
rational term and obtain 
the exact solution in terms of $X_1$ exceptional Laguerre polynomials. The 
generalization to the $X_m$ Laguerre case is discussed in subsection $3.1$. 
Finally we summarize our results in section $4$.


\section{The conventional TCS model}
The $N$-body TCS model \cite{pitt_17} is characterized by the Hamiltonian
\ba\label{tcm}
\hat{H}=\sum^{N}_{i=1}\bigg[-\frac{1}{2}\frac{\partial^2}{\partial x^2_i}+\frac{1}{2}\omega^2x^2_i\bigg]+V_{int},
\ea
where  
\be\label{vi}
V_{int}=\sum_{\substack {i<j \\ \mid i-j\mid \le r}}\frac{\lambda (\lambda-1)}{\mid x_i-x_j\mid ^2}+\sum_{\substack {i<j<k \\ \mid i-j\mid \le r\\ \mid j-k\mid \le r}}\frac{\lambda^2 {\bf r}_{ij}.{\bf r_{jk} }}{r^2_{ji}r^2_{jk}};\qquad \lambda \ne 0,
\ee 
i.e. the particles are interacting through a pair wise two body potential as
well as a three body term. The vector along 
$x$-axis is ${\bf r}_{ij}=(x_i-x_j)\hat x$. The above two body interactions are attractive for $0<\lambda<1$
and repulsive for $\lambda\ge 1$. It is worth pointing out that in the
particular cases of $r = 1$ and $r = N-1$, this Hamiltonian reduces to
those of JK \cite{jain_khare_99} and CSM \cite{cal_71,suth_71} respectively.\\  
The solution of the above model is obtained in \cite{pitt_17} and is given by
\be\label{tcsm_sol}
\Psi({\bf x})=\phi({\bf x})\xi({\bf x}); \qquad {\bf x}=(x_1,x_2,....,x_N) \epsilon \mathbb{R}^N,
\ee 
where 
\be\label{phi}
\phi({\bf x})=\prod_{i<j} (x_i-x_j)^{\lambda}\,
\ee
while the function $\xi$ satisfies the equation 
\be\label{uxi}
-\frac{1}{2}\sum^{N}_{i=1}\frac{\partial^2 \xi}{\partial x^2_i}-\lambda\sum^{N-1}_{i<j}\frac{1}{x_i-x_j}\bigg(\frac{\partial \xi}{\partial x_i}-\frac{\partial \xi}{\partial x_j}\bigg)+\bigg(\frac{1}{2}\sum^N_i\omega^2x^2_i-E\bigg)\xi=0.
\ee
To get exact solutions of the above equation, one assume $\xi$ as
\be\label{xs}
\xi=\Phi(\rho) P_s({\bf x}); \qquad \mbox{where}\quad \rho^2=\sum^N_{i=1} x^2_i.
\ee 
Substituting  $\xi({\bf x})$ in Eq. (\ref{uxi}), one finds that $\Phi$
satisfies the differential equation 
\be\label{srho}
\Phi''(\rho)+\bigg(N+2s-1+ \lambda r(2N-r-1)\bigg)\frac{1}{\rho}\Phi'(\rho)
+2(E-\frac{1}{2}\omega^2\rho^2)\Phi(\rho)=0\,,
\ee
while the function $P_s({\bf x})$ behaves as a homogeneous polynomial of 
degree $s (=0,1,2,...)$ and satisfies a generalized Laplace equation
\be\label{ps}
\bigg[\sum^{N}_{i=1}\frac{\partial^2 }{\partial x^2_i}+2\lambda\sum^{N-1}_{i<j}\frac{1}{x_i-x_j}\bigg(\frac{\partial }{\partial x_i}-\frac{\partial}{\partial x_j}\bigg) \bigg ] P_s({\bf x})=0.
\ee
The solutions of this Laplace equation are discussed in detail in 
Refs. \cite{ajk_01,jain_khare_99, pitt_17} and the 
 Eq. (\ref{srho}) is the well known equivalent radial equation for the 
oscillator potential in  arbitrary dimensions. 
The solution of this radial equation is in terms of the classical Laguerre 
orthogonal polynomial $(L^{(\alpha)}_n(\omega \rho^2))$
and is given by 
\be
\Phi(\rho)\simeq \exp(-\frac{\omega \rho^2}{2})L^{(\alpha)}_n(\omega \rho^2 );\quad n=0,1,2,...
\ee
while the corresponding energy eigenvalues are 
\be\label{con_en}
E_n=\omega \big (2n+s+\frac{N}{2}+\frac{\lambda r}{2} (2N-r-1)\big),
\ee
where $\alpha=\big( s-1+\frac{N}{2}+\frac{\lambda r}{2} (2N-r-1)\big)$. 
As shown in \cite{pitt_17}, for $r=1$ and $r = N-1$, the results 
reduces to those of JK 
\cite{jain_khare_99} and CSM \cite{cal_71,suth_71} respectively.

\section{The extended truncated CS model with new interaction term}
The above $N$-body truncated CS model can be extended by adding a new interaction term $V_{new}$ as 
\ba\label{etcm}
\hat{H}_{ext}=\hat{H}+V_{new},
\ea
where 
\be\label{vnew}
V_{new}=\frac{(\alpha_1+\alpha_2 \omega^2 \rho^2)}{(\beta_1+\beta_2 \omega^2 \rho^2)^2},
\ee 
 where $\alpha_{1,2}$  and $\beta_{1,2}$ are unknown constants. Following the procedure adopted above in the case 
of conventional model, the solution of the Schr\"odinger equation 
\be
\hat{H}_{ext}\Psi_{ext}=E_{ext}\Psi_{ext}
\ee
corresponding to the extended Hamiltonian ($\hat{H}_{ext}$) is obtained by assuming the extended wavefunction 
\be\label{extcsm_sol}
\Psi_{ext}({\bf x})=\phi({\bf x})\xi_{ext},
\ee 
where $\phi({\bf x})$ again is as given by Eq. (\ref{phi}) while 
 $\xi_{ext}$ satisfies the equation 
\be\label{euxi}
-\frac{1}{2}\sum^{N}_{i=1}\frac{\partial^2 \xi_{ext}}{\partial x^2_i}-\lambda\sum^{N-1}_{i<j}\frac{1}{x_i-x_j}\bigg(\frac{\partial \xi_{ext}}{\partial x_i}-\frac{\partial \xi_{ext}}{\partial x_j}\bigg)+\bigg(\frac{1}{2}\sum^N_i\omega^2x^2_i+V_{new}-E_{ext}\bigg)\xi_{ext}=0.
\ee
As in the conventional case as discussed above, we redefine the function 
$\xi_{ext}$ as
\be\label{exs}
\xi_{ext}=\Phi_{ext}(\rho) P_s({\bf x}).
\ee 
In that cas Eq. (\ref{euxi}) reduces to the $\rho$ dependent equation  
\be\label{esrho}
\Phi_{ext}''(\rho)+\big(N+2s-1+ \lambda r(2N-r-1)\big)\frac{1}{\rho}\Phi_{ext}'(\rho)+2\big(E-(\frac{1}{2}\omega^2\rho^2+V_{new})\big )\Phi_{ext}(\rho)=0,
\ee
with $P_s({\bf x})$ satisfying  the same generalized Laplace equation 
Eq. (\ref {ps}). Note that here a prime on $\Phi_{ext}(\rho)$ indicates 
derivative with respect to $\rho$.\\
To get the exact form of the defined new interaction term (\ref{vnew}) and the solutions of the above equation,
 we assume 
\be\label{extsol}
\Phi_{ext}(\rho)=f(\rho)\zeta(g(\rho)),
\ee 
where $f(\rho)$ and $g(\rho)$ are two undermined 
functions and $\zeta(g)$ is a special function 
which satisfies a second-order differential equation
\be\label{de}
\zeta''(g(\rho))+Q_1(g)\zeta'(g(\rho))+R_1(g)\zeta(g(\rho))=0.
\ee
The functions $Q_1(g)$ and $R_1(g)$ are well defined for any special function $\zeta(g)$.
Substituting Eq. (\ref{extsol}) into Eq. (\ref{esrho}), we get 
\ba\label{ede}
\zeta''(g)&+&\bigg(\frac{2f'(\rho)}{f(\rho )g'(\rho)}+\frac{g''(\rho)}{g'(\rho)^2}+\frac{\tau}{\rho g'(\rho)}\bigg)\zeta'(g)\nonumber\\
&+&\frac{1}{g'(\rho)^2}\bigg( \frac{f''(\rho)}{f(\rho)}+\frac{\tau f'(\rho)}{\rho f(\rho)}+2(E_{ext}-V_{ext})\bigg)\zeta(g)=0,
\ea
where $V_{ext}=\frac{1}{2}\omega \rho^2+V_{new}$ and $\tau=\big(N+2s-1+ \lambda r(2N-r-1)\big)$.
On comparing Eq. (\ref{ede}) with Eq. (\ref{de}), we get
\ba\label{q}
Q_1(g)&=&\frac{2f'(\rho)}{f(\rho )g'(\rho)}+\frac{g''(\rho)}{g'(\rho)^2}+\frac{\tau}{\rho g'(\rho)}\\
\mbox{and}\quad R_1(g)&=& \frac{1}{g'(\rho)^2}\bigg( \frac{f''(\rho)}{f(\rho)}+\frac{\tau f'(\rho)}{\rho f(\rho)}+2(E_{ext}-V_{ext})\bigg).
\ea
After simplifying $Q_1(g)$, one finds that
\be\label{f}
f(\rho)\simeq (g'(\rho))^{-\frac{1}{2}}\rho^{-\frac{\alpha}{2}}\exp\bigg(\frac{1}{2}\int^{g}Q_1(g)dg\bigg).
\ee  
Using $f(\rho)$ in the expression of $R_1(g)$ we get 
\be\label{e_v}
E_{ext}-V_{ext}=\frac{1}{2}\bigg[ \frac{g'''(\rho)}{2g'(\rho )}-\frac{3}{4}\frac{g''(\rho)^2}{g(\rho)^2}+\frac{\tau/2(\tau/2-1)}{\rho^2} + g'(\rho)^2\bigg(R_1(g)-\frac{Q_1'(g)}{2} - \frac{Q_1^2(g)}{4}\bigg)\bigg].
\ee
Thus, once we choose $Q_1(g)$ and $R_1(g)$ corresponding to the given special function $\zeta(g)$ the extended potential $V_{ext}$ and 
the corresponding energy $E_{ext}$  can be obtained for given $g(\rho)$ as 
defined in the case of conventional model.

Let us consider the special function $\zeta(g)$  in the form of $X_1$ Laguerre 
polynomial $\hat{L}^{(\alpha)}_{n}(g)$ satisfying the differential equation
\be\label{ex}
\hat{L}^{''(\alpha)}_{n}(g(\rho))+Q(g)\hat{L}^{'(\alpha)}_{n}(g(\rho))
+R(g)\hat{L}^{(\alpha)}_{n}(g(\rho))=0;\qquad n\ge 1,
\ee
with 
\ba\label{qg}
Q_1(g)&=&-\frac{(g-\alpha )( g+\alpha+1 )}{g(g+\alpha)} \nonumber\\ 
\mbox{and}\quad R_1(g)&=& \frac{1}{g}\bigg ( \frac{(g-\alpha )}{(g+\alpha)}+n-1\bigg).
\ea
Using above equations in Eqs. (\ref{f}) and (\ref{e_v}) and by defining
\be 
g(\rho)=\omega \rho^2; \quad \alpha=\frac{\tau}{2}-\frac{1}{2}
\ee
 and replacing $n\rightarrow n+1$, we get
\ba\label{ext_new}
V_{ext}=\frac{1}{2}\omega^2 \rho^2+\frac{4\omega}{(2\omega \rho^2+\tau-1)}-\frac{8\omega(\tau-1)}{(2\omega \rho^2+\tau-1)^2},
\ea
and the energy eigenvalues $E_{ext}$ turn out to be the same as that of the 
conventional model as discussed in Sec. II and are given by Eq. (\ref{con_en}).
 Note however that the corresponding eigenfunction $\Phi_{ext}(\rho)$ is 
completely different. Using $f(\rho)$ and replacing 
$\zeta(g)\rightarrow \hat{L}^{(\alpha)}_{n+1}(g)$ in Eq. (\ref{extsol}), the 
expression for the energy eigenfunctions is obtained in terms of $X_1$ 
exceptional orthogonal Laguerre polynomials $(\hat {L}^{(\alpha)}_{n+1}(g))$ as
\be\label{x_1}
\Phi_{ext}(\rho)\simeq \frac{\exp(-\frac{\omega \rho^2}{2})}{(2\omega \rho^2+\alpha)}\hat {L}^{(\alpha)}_{n+1}(\omega \rho^2 );\quad n=0,1,2,...,.
\ee 
Note that the $X_1$ Laguerre 
polynomial $(\hat {L}^{(\alpha)}_{n+1}(g))$ is related to the classical 
Laguerre polynomials by 
\be
\hat{L}^{(\alpha)}_{n+1} (g) = -(g + \alpha + 1)L^{(\alpha)}_n(g)+ L^{(\alpha)}_{n-1}(g).
\ee  
The constant parameters $\alpha_{1,2}$ and $\beta_{1,2}$ for which the 
Hamiltonian (\ref{tcm}) is ES can easily be determined by comparing 
Eqs. (\ref{etcm})  and (\ref{ext_new}) and one finds that
\ba\label{ab}
\alpha_1&=&-4\omega(\tau-1);\qquad \alpha_2=8,\nonumber \\
\beta_1&=&\tau-1; \quad \mbox{and}\quad \beta_2=2/\omega.
\ea
In the special cases of $r = 1$ and $r = N-1$ we then obtain the rational 
extension of the JK and the CSM respectively.
\subsection{The extended TCS model associated with $X_m$-exceptional Laguerre 
polynomials} 

The above model Eq. (\ref{etcm}) can easily be generalized to any positive 
integer values of $m$ by replacing $V_{new}$ with an $m$ dependent polynomial 
type interaction term  $V_{m, new}$ i.e., 
\ba\label{etcsm}
\hat{H}_{m, ext}=\hat{H}+V_{m,new}.
\ea
Unlike the $X_1$ case, it is not easy to define the exact form of $V_{m,new}$ 
in the general $X_m$ case. We shall obtain the interaction terms by assuming 
the solution of the Schr\"odinger equation 
\be
\hat{H}_{m,ext}\Psi_{ext}({\bf x})=E_{m, ext}\Psi_{m, ext}({\bf x})
\ee
as
\be\label{extcsm_msol}
\Psi_{m,ext}({\bf x})=\phi({\bf x})\xi_{m,ext}({\bf x}),
\ee 
Similar to the $X_1$ case, we redefine Eqs. (\ref{exs}) and (\ref{extsol}) by replacing $\Phi_{ext}(\rho)\rightarrow \Phi_{m,ext}(\rho)$ and 
$f(\rho)\rightarrow f_m(\rho), \zeta(g)\rightarrow \zeta_m(g)$ respectively. 
In this way the differential Eq. (\ref{esrho}) will be also $m$ dependent i.e.,   
\be\label{esrho_m}
\Phi_{m,ext}''(\rho)+\big(N+2s-1+ \lambda r(2N-r-1)\big)\frac{1}{\rho}\Phi_{m,ext}'(\rho)+2\big(E-(\frac{1}{2}\omega^2\rho^2+V_{m,new})\big )\Phi_{m,ext}(\rho)=0.
\ee
 and $\zeta_m(g)$ satisfies an equivalent second-order differential equation
\be\label{de_m}
\zeta_{m}''(g)+Q_m(g)\zeta'_m(g)+R_m(g)\zeta_m(g)=0.
\ee
Now using $\Phi_{m,ext}(\rho)$ into Eq. (\ref{esrho_m}), we get 
\ba\label{ede_m}
\zeta_m''(g)&+&\bigg(\frac{2f_m'(\rho)}{f_m(\rho )g'(\rho)}+\frac{g''(\rho)}{g'(\rho)^2}+\frac{\tau}{\rho g'(\rho)}\bigg)\zeta_m'(g)\nonumber\\
&+&\frac{1}{g'(\rho)^2}\bigg( \frac{f_m''(\rho)}{f_m(\rho)}+\frac{\tau f'(\rho)}{\rho f(\rho)}+2(E_{m,ext}-V_{m,ext})\bigg)\zeta_m(g)=0,
\ea
where $V_{m,ext}=\frac{1}{2}\omega \rho^2+V_{m,new}$. The functions $Q_m(g)$ i
and $R_m(g)$ become
\ba\label{q_m}
Q_m(g)&=&\frac{2f_m'(\rho)}{f_m(\rho )g'(\rho)}+\frac{g''(\rho)}{g'(\rho)^2}+\frac{\tau}{\rho g'(\rho)}\\
\mbox{and}\quad R_m(g)&=& \frac{1}{g'(\rho)^2}\bigg( \frac{f_m''(\rho)}{f_m(\rho)}+\frac{\tau f'(\rho)}{\rho f_m(\rho)}+2(E_{m,ext}-V_{m,ext})\bigg).
\ea
In terms of $Q_m(g)$, the function $f_m(\rho)$ is given by 
\be\label{f_m}
f_m(\rho)\simeq (g'(\rho))^{-\frac{1}{2}}\rho^{-\frac{\alpha}{2}}\exp\bigg(\frac{1}{2}\int^{g}Q_m(g)dg\bigg).
\ee  
Using $f_m(\rho)$ back in the expression of $R_m(g)$ and get 
\be\label{ev_m}
E_{m,ext}-V_{m,ext}=\frac{1}{2}\bigg[ \frac{g'''(\rho)}{2g'(\rho )}-\frac{3}{4}\frac{g''(\rho)^2}{g(\rho)^2}+\frac{\tau/2(\tau/2-1)}{\rho^2} + g'(\rho)^2\bigg(R_m(g)-\frac{Q_m'(g)}{2} - \frac{Q_m^2(g)}{4}\bigg)\bigg].
\ee
Similar to the $X_1$ case, the special function $\zeta_m(g)$ satisfies 
the $X_m$ exceptional Laguerre differential \cite{xm1,xm2,dim}
\be\label{ex_m}
\hat{L}^{''(\alpha)}_{n,m}(g(\rho))+Q_m(g)\hat{L}^{'(\alpha)}_{n,m}(g(\rho))
+R_m(g)\hat{L}^{(\alpha)}_{n,m}(g(\rho))=0,
\ee
with 
\ba\label{qg_m}
Q_m(g)&=&\frac{1}{g}\bigg[(\alpha+1-g)-2g\frac{L^{(\alpha)}_{m-1}(-g)}{L^{(\alpha-1)}_{m}(-g)}\bigg] \nonumber\\ 
\mbox{and}\quad R_m(g)&=& \frac{1}{g}\bigg[n-2\alpha\frac{L^{(\alpha)}_{m-1}(-g)}{L^{(\alpha)}_{m}(-g)}\bigg].
\ea
Using above $Q_m(g)$ and $R_m(g)$ in Eqs. (\ref{f_m}) and (\ref{ev_m}) and replacing $n\rightarrow n+m$, we get
\ba\label{ext_new_m}
V_{m,new}&=&-2\omega^2\rho^2\frac{L^{(\alpha+1)}_{m-2}(-g)}{L^{(\alpha-1)}_{m}(-g}+2\omega(\alpha+\omega \rho^2-1)\frac{L^{(\alpha)}_{m-1}(-g)}{L^{(\alpha-1)}_{m}(-g}\nonumber\\
&+&4\omega^2\rho^2\bigg(\frac{L^{(\alpha)}_{m-1}(-g)}{L^{(\alpha-1)}_{m}(-g}\bigg)^2-2m\omega,
\ea
while the energy eigenvalues $E_{m,ext}$ are again unchanged and are same as that of the $X_1$ or conventional cases i.e, $E_{m,ext}=E_{ext}=E$.
The energy eigen functions  $\Phi_{m,ext}(\rho)$ are however different and
are given by 
\be\label{x_m}
\Phi_{m,ext}(\rho)\simeq \frac{\exp(-\frac{\omega \rho^2}{2})}{\hat{L}^{(\alpha-1)}_m(-\omega\rho^2)}\hat {L}^{(\alpha)}_{n+m}(\omega \rho^2 );\quad n,m=0,1,2,....,
\ee 
where the $X_m$ Laguerre polynomial $(\hat {L}^{(\alpha)}_{n+m}(g))$ is
related to the classical Laguerre polynomials by
\be
\hat{L}^{(\alpha)}_{n+m} (g) = L^{(\alpha)}_m(-g)L^{(\alpha-1)}_n(g)+ L^{(\alpha-1)}_m(-g)L^{(\alpha)}_{n-1}(g).
\ee 
As expected, for $m=1$, the above results reduce to the corresponding 
$X_1$-case while for the $m=0$ case one gets back the conventional TCS model.
\section{Results and discussion}
In this paper we have constructed an extended truncated Calogero-Sutherland 
model by introducing new interaction terms. The 
exact solutions of this extended model are in terms of the newly discovered 
special function, the $X_1$-exceptional Laguerre Polynomials while the energy
eigenvalues remain unchanged and are same as those of TCS model. The model 
is further extended to the $X_m$ case and the corresponding 
$m$-dependent interaction term is obtained. In the particular case of $m=0$ 
and $r=1$ or $r = N-1$, it can be easily shown that the 
Hamiltonian (\ref{etcsm}) and the corresponding eigenvalues and eigenfunctions 
reduce to that of JK model 
 or CSM respectively. Thus for $r =1$ or $r = N-1$, 
one obtains extended JK model or CSM corresponding to the $X_m$-case 
simply by putting $r=1$ or $r = N-1$ in Eq. (\ref{etcsm}). 

This paper raises some obvious possibilities. What we have done in this
paper is basically obtained rational extension of $A_N$ JK or TCS models.
The obvious question is can one extend these results to the other cases
like $B_N, C_N, BC_N, D_N$ or even to the exceptional groups? Further,
are there other N-body problems where such rational extensions are possible?
We hope to address some of these issues in the near future.

{\bf Acknowledgments}\\
 A.K. wishes to thank Indian National Science Academy (INSA) for 
the award of INSA senior scientist position at Savitribai Phule Pune University.


\begin{thebibliography}{99}
\bibitem{dnr1} D. Gomez-Ullate, N. Kamran and R. Milson, {\it J. Math. Anal.Appl.} \textbf{359} (2009) 352.  
\bibitem{xm1} D. Gomez-Ullate, N. Kamran and R. Milson, {\it J. Phys. A}  43 (2010) 434016.  
\bibitem{xm2} D. Gomez-Ullate, N. Kamran and R. Milson, {\it Contemporary Mathematics} {\bf 563}  (2012) 51.
\bibitem{que}  C. Quesne, {\it J.Phys.A} \textbf{41} (2008) 392001.
\bibitem{bqr}  B. Bagchi, C. Quesne and R. Roychoudhary, 
{\it Pramana J. Phys.} \textbf{73}(2009) 337, C. Quesne, SIGMA {\bf 5} (2009)
84.
\bibitem{os}  S. Odake and R. Sasaki, {\it Phys. Lett. B}, \textbf{684} 
(2010) 173; ibid {\bf 679} (2009) 414. {\it J. Math. Phys}, \textbf{51}, (2010) 053513.
\bibitem{hos} C-L. Ho, S ODAKE and R Sasaki, {\it SIGMA} \textbf{7} (2011) 107. 
\bibitem{hs} C-L. Ho and R Sasaki, {\it ISRN Math. Phys.} 2012 (2012) 920475. 
\bibitem{gom} D. Gomez-Ullate, N. Kamran and R. Milson, {\it J. Math. Anal. Appl.} 399(2) (2013) 480.
\bibitem{qu} C. Quesne, {\it Int. J. Mod. Phys. A} \textbf{26} (2011) 5337.
\bibitem{yg1} Y. Grandati, {\it Ann. Phys.} \textbf{326} (2011) 2074; \textbf{327} (2012) 2411.
\bibitem{dim} R. K. Yadav et al., {\it Acta Polytechnica} \textbf{57} (2017) 477. 



\bibitem{pdm} B. Midya and B. Roy, {\it Phys. Lett. A} \textbf{373} (2009) 4117.
\bibitem{nfold2} B. Midya, B. Roy, and T. Tanaka, {\it J. Phys. A} \textbf{45} (2012) 205303.
\bibitem{qscat}  R. K. Yadav, A. Khare and B. P. Mandal, {\it Annals of Physics} \textbf {331} (2013) 313; {\it Phys. Lett. B} \textbf{723} (2013) 433; {\it Phys. Lett. A} \textbf{379} (2015) 67.
\bibitem{scatpt}  N. Kumari, R. K. Yadav, A. Khare, B. Bagchi, B. P. Mandal,
{\it Annals of Physics} {\bf 373} (2016) 163.

\bibitem{tdse} A. Schulze-Halberg and B. Roy, {\it J. Math. Phys.} \textbf{55} (2014) 123506.

\bibitem{gtextd}  R. K. Yadav, N. Kumari, A. Khare and B. P. Mandal, {\it Annals of Physics} \textbf {359} (2015) 46.
\bibitem{gtextd2}  A Ramos et al., {\it Annals of Physics} \textbf {382} (2017) 143.
\bibitem{para_sym} R. K. Yadav, A. Khare, B. Bagchi,  N. Kumari, B. P. Mandal,
{\it J. Math. Phys.} {\bf 57} (2016) 062106-1.
\bibitem{nrab} N. Kumari, R. K. Yadav, A. Khare and B P Mandal,{\it Annals of Physics} \textbf {385} (2017) 57. 
\bibitem{nrab_nc} N. Kumari, R. K. Yadav, A. Khare and B P Mandal,{\it J. Math. Phys.} \textbf {59} (2018) (062103-1). 
\bibitem{cal_71} F. Calogero, {\it J. Math. Phys.} \textbf {12} (1971) 419.
\bibitem{suth_71} B. Sutherland, {\it J. Math. Phys.} \textbf {12} (1971) 246.
\bibitem{bpm5} B Basu-Mallick, B P Mandal and P Roy, {\it Annals of Physics} \textbf {380} (2017) 206.
\bibitem{ajk_01} Guy Anberson, S. R. Jain and A. Khare, {\it J. Phys. A} \textbf {34} (2001) 695. 
\bibitem{jain_khare_99} S. R. Jain and A. Khare, {\it Phys. Lett. A } \textbf {262} (1999) 35.
\bibitem{pitt_17} S. M. Pittman, M. Beau, M. Olshanii and A. del Campo, {\it Phys. Rev. B } \textbf {95} (2017) 205135.



 
              

 
                 






\end{thebibliography}
\end{document}